\documentclass[aps,prd,twocolumn,showpacs,preprintnumbers,superscriptaddress,nofootinbib]{revtex4}
\usepackage{graphics}  
\usepackage{epsfig}
 \usepackage{verbatim}

\IfFileExists{srcltx.sty}{\usepackage[active]{srcltx}}

\newcommand{\be}{\begin{equation}}
\newcommand{\ee}{\end{equation}}
\newcommand{\bea}{\begin{eqnarray}}
\newcommand{\eea}{\end{eqnarray}}

\begin{document}

\title{Role of Galactic sources and magnetic fields in forming the observed energy-dependent composition of ultrahigh-energy cosmic rays}

\author{Antoine  Calvez}
\affiliation{Department of Physics and Astronomy, University of California, Los
Angeles, CA 90095-1547, USA}
\author{Alexander Kusenko}
\affiliation{Department of Physics and Astronomy, University of California, Los
Angeles, CA 90095-1547, USA}
\affiliation{IPMU, University of Tokyo, Kashiwa, Chiba 277-8568, Japan}
 \author{Shigehiro Nagataki}
\affiliation{Yukawa Institute for Theoretical Physics, Kyoto University, Oiwakecho, Kitashirakawa, Sakyoku, Kyoto
606-8502, Japan}



\begin{abstract}
Recent results from the Pierre Auger Observatory, showing energy-dependent chemical composition of ultrahigh-energy cosmic rays (UHECRs) with a growing fraction of heavy elements at high energies, suggest a possible non-negligible contribution of the Galactic sources.  We show that, in the case of UHECRs produced by gamma-ray bursts (GRBs) or rare types of supernova explosions that took place in the Milky Way in the past, the change in UHECR composition can result from the difference in diffusion times for different species.  The anisotropy in the direction of the Galactic Center is expected to be a few per cent on average, but the locations of the most recent/closest bursts can be associated with observed clusters of UHECRs.
\end{abstract}

\maketitle

Recent measurements of the composition of ultrahigh-energy cosmic rays (UHECRs) by the Pierre Auger Observatory (PAO) have
suggested that the mean nuclear mass may increase with energy
between 2~EeV and 35~EeV~\cite{Abraham:2009dsa}. This was not expected. The lack of plausible sources in the Milky Way (MW) and the lack of Galactocentric anisotropy of the arrival directions of UHECRs are usually seen as evidence for extragalactic origin of UHECRs above 10$^{18}$~eV.

However, if the cosmic rays of 10$^{18}-10^{19}$~eV are nuclei, the turbulent Galactic micro-Gauss magnetic fields~\cite{MF} can retain them in the Galaxy and isotropize their directions sufficiently to show no disagreement with the data.  Moreover, since the diffusion times depend on the rigidity, the observed composition can be altered by diffusion~\cite{Wick:2003ex}.
Since the heavier nuclei spend more time in the Galaxy than the lighter nuclei and protons, they have the higher number density and flux. Thus, diffusion alone can alter the composition of UHECRs produced by the Galactic sources and increase the observed fraction of nuclei.

As for the plausible sources, there is a growing evidence that long GRBs are caused by a relatively rare type of supernovae, while the short GRBs probably result from the coalescence of neutron stars and/or black holes~\cite{shGRBs_from_compact_stars}. Compact star mergers undoubtedly take place in the Milky Way, and therefore short GRBs should occur in our Galaxy. Although there is some correlation of long GRBs with star-forming metal-poor galaxies~\cite{Fruchter:2006py}, many long GRBs are observed in high-metallicity galaxies as well~\cite{GRBs_hi_m_galaxies}, and therefore one expects that long GRBs should occur in the Milky Way.  Less powerful hypernovae, too weak to produce a GRB, but can still accelerate UHECRs~\cite{Wang:2007ya}, with a substantial fraction of nuclei~\cite{Wang:2007xj_nuclei_survival}.

We will show that the observed change in chemical composition can be explained, at least qualitatively, by the past Galactic GRBs producing UHECRs that diffuse in the $\sim 3 \mu$G magnetic field.  Changes in composition due to a magnetic fields have  been discussed in connection with the spectral ``knee''~\cite{Wick:2003ex}, and also for a transient source~\cite{Kotera:2009ms}. We illustrate the effect of Galactic diffusion using a simple model and show that, in some range of energies, the {\em observed} composition is energy-dependent even if the spectrum {\em produced} at the source does not have an energy-dependent composition.  Moreover, for reasonable values of the Galactic magnetic fields, the transition to
energy-dependent composition is expected around $\sim 10^{18}$~eV, in agreement with the data.  We  concentrate on UHECRs with the energies below the Greisen-Zatsepin-Kuzmin (GZK) cutoff~\cite{gzk}.  We note that the PAO results on chemical composition~\cite{Abraham:2009dsa} are in agreement with the Yakutsk experiment~\cite{Glushkov:2007gd}, but they have not been corroborated by the HiRes experiment~\cite{hires}, and there remains a significant experimental uncertainty due to possible systematic errors.

GRBs have been proposed as the sources of extragalactic UHECRs~\cite{Waxman:1995vg}, and they have also been considered as possible Galactic sources~\cite{Dermer:2005uk}. It is believed that GRBs, hypernovae, or other stellar events capable of producing UHECRs could have happened in the Milky Way at the rate of one per  $t_{GRB}\sim 10^{4}-10^{5}$ years \cite{Schmidt:1999iw}. Such events have been linked to the observations of INTEGRAL, Fermi and PAMELA~\cite{Bertone:2004ek}.

Let us consider a simple model of cosmic ray transport. The spectrum of UHECRs is assumed to contain different nuclei, which we will label by their electric charges $q_i=e Z_i$.  Let us denote the number density of $i$'th species of cosmic rays with energy $E$ at point $\vec{r}$ by $n_i(E,\vec{r},t)$.  We assume for simplicity that production of cosmic rays of different species occurs in such a way that the production spectra $Q_i(E,\vec{r})$ have the same energy dependence, and that the chemical composition is energy independent, given by the constants $\xi_i$:
\begin{equation}
Q_i(E,\vec{r})= Q_0 \xi_i (E/E_0)^{-\gamma} \rho(\vec{r}),
\label{production}
\end{equation}
where $Q_0$ and $E_0$ do not depend on the type of the species or the energy.  We will consider three different models for the distribution of sources $\rho(\vec{r})$.

We now ask whether the {\em measured} fluxes inside the galaxy can be altered by diffusion. In diffusive approximation, the transport inside the galaxy can be described by the usual equation:
\begin{eqnarray}
\frac{\partial n_i}{\partial t} - \vec{\nabla}(D_i \vec{\nabla}n_i)  +\frac{\partial}{\partial E}(b_i n_i )= \nonumber \\
 Q_i(E,\vec{r},t) +\sum_k\int P_{ik}(E,E') n_k(E') dE'.
\label{transport}
\end{eqnarray}
Here $D_i(E,\vec{r},t)=D_i(E)$ is the diffusion coefficient, which we will assume to be constant in space and time. The energy losses and all the interactions that change the particle energies
are given by $b_i(E)$ and the kernel in the collision integral $P_{ik}(E,E')$. For energies below GZK cutoff, one can neglect the energy losses on the diffusion time scales.

The diffusion coefficient $D(E)$ depends primarily on the structure of the magnetic fields in the galaxy. Let us assume that the magnetic field structure is comprised of uniform randomly oriented domains of radius $l_0$ with a constant field $B$ in each domain. The density of such domains is $N\lesssim l_0^{-3}$. The Larmor radius depends on the particle energy $E$ and its electric charge $q_i=eZ_i$:
\begin{eqnarray}
R_i & =& \frac{E}{B q_i} = l_0 \left( \frac{E}{E_{0,i}} \right),\ {\rm where} \
E_{0,i}= E_0 \, Z_i, \\
E_0 & = & B l_0 = 10^{18} {\rm eV} \left(\frac{B}{3\times 10^{-6}\, {\rm G}}\right) \left(\frac{l_0}{0.3\, {\rm kpc}} \right)   . \label{E0i}
\end{eqnarray}

The spatial energy spectrum of random magnetic fields inferred from observations suggests that $B\sim 3\mu {\rm G} $ on the 0.3~kpc spatial scales, and that there is a significant change at $l=1/k\sim 0.1-0.5$~kpc~\cite{MF}. This can be understood theoretically because the turbulent energy is injected into the interstellar medium by supernova explosions on the scales of order 0.1~kpc.  This energy is transferred to smaller scales by direct cascade, and to larger scales by inverse cascade of magnetic helicity.  Single-cell-size models favor $\sim 0.1$~kpc scales as well~\cite{MF}.

For each species, there is a critical energy $E_{0,i}$ for which the Larmor radius is equal to the magnetic coherence length $l_0$. In this simple model, the diffusion proceeds in two very different regimes.  For $E\lesssim E_{0,i}$, the mean free path of the diffusing particle is $l\sim 1/(N l_0^2) \sim l_0$, and, in the units for which the speed of light $c=1$, the diffusion coefficient is
\begin{eqnarray}
D_i(E)= \frac{1}{3}l_0 \equiv D_0, \ {\rm for} \ E \ll  E_{0,i}.
\end{eqnarray}

\begin{figure}
\begin{center}
 \includegraphics[width=0.99 \columnwidth]{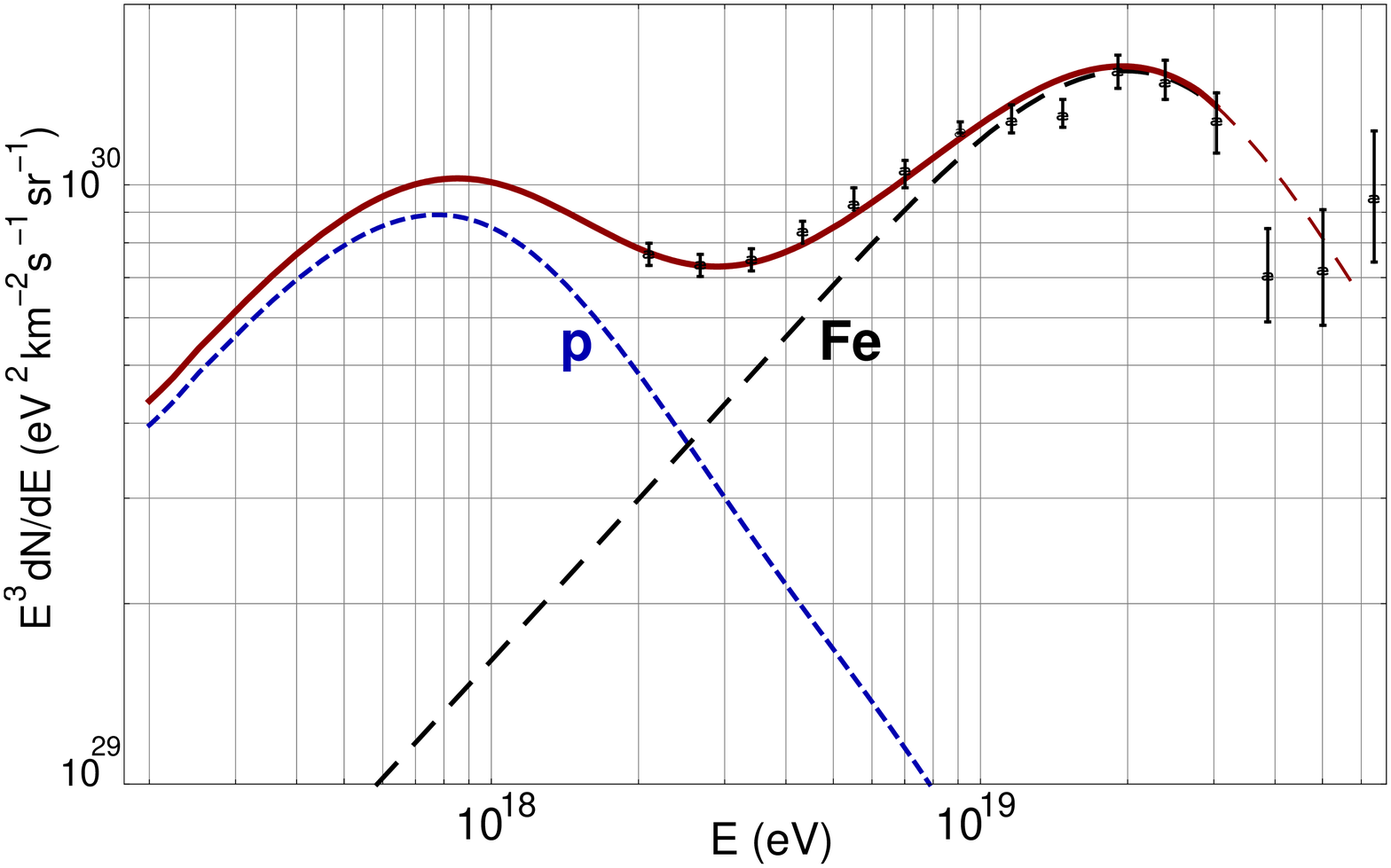}
\caption{Predicted UHECR spectra, assuming that the sources produce 90\% protons and 10\% iron, with identical spectra {$\propto  E^{-2.3}$}, and that the source distribution traces the distribution of stars in the Galaxy.  We used samples of $10^3$ GRBs at random locations with time intervals of $10^5$ years.  The magnetic field was assumed to be $4~\mu$G, coherent over $l_0=0.2$~kpc domains, $\delta_1=0.3$, $\delta_2=0$. The overall power and the iron fraction were adjusted to fit the PAO data points~\cite{Abraham:2009wk} (shown). For each random sample, the fit parameters differ slightly, depending on the location of the latest/closest burst.  To model the spectrum at $E>3\times 10^{19}$~eV, one has to account for energy losses
and (proton) contribution of extragalactic sources, which we leave for future work. }
    \end{center}
  \label{fig:spectrum_distr}
\end{figure}

\begin{figure}
\begin{center}
 \includegraphics[width=0.99 \columnwidth]{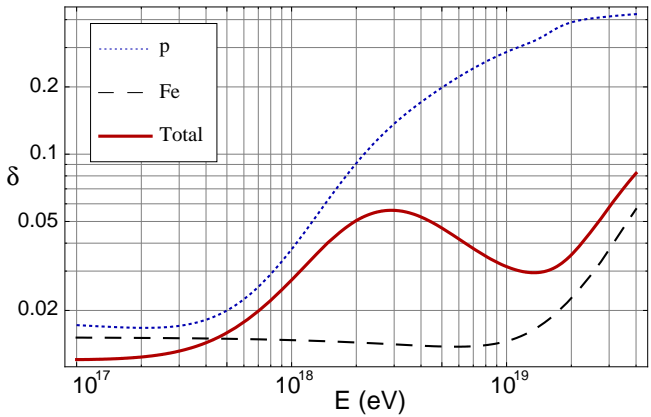}
\caption{Galactocentric anisotropy for a source distribution that traces the stellar counts in MW, modeled by random generation of $10^3$ bursts separated by time intervals of $10^5$yr. The model parameters are the same as in Fig.~1. Although the anisotropy in protons is large at high energies, their contribution to the total flux is small, so the total anisotropy $<10\%$, consistent with the observations. The latest GRBs do not introduce a large degree of anisotropy, as it would be in the case of UHE protons, but they can create ``hot spots'' and clusters of events.
  }
    \end{center}
  \label{fig:anisotropy_stars}
\end{figure}

For $E\gg E_{0,i}$, the particle is deflected only by a small angle $\theta\sim l_0/R_i$, and, after $k$ deflections, the mean deflection angle squared is $\bar{\theta^2}\sim k (l_0/R_i)^2$.  The corresponding diffusion coefficient is
\begin{eqnarray}
 D_i(E) = D_0 \left ( \frac{E}{E_{0,i}} \right)^2 , \ {\rm for} \ E \gg  E_{0,i}.
\end{eqnarray}
While this simple model is reasonable and well known~\cite{Ginzburg}, in reality the energy dependence of the diffusion coefficient can be more complicated, for example, because the magnetic field domains are different in size~\cite{MF}.  To account for this, we introduce two parameters $0 \le \delta_{1,2} \le 0.5$ and assume the following behavior of the diffusion coefficient:
\begin{equation}
 D_i(E) = \left \{
\begin{array}{ll}
D_0 \left ( \frac{E}{E_{0,i}} \right)^{\delta_1}, & E \le  E_{0,i}, \\
D_0 \left ( \frac{E}{E_{0,i}} \right)^{(2-\delta_2)},  & E > E_{0,i} .
\end{array} \right.
\end{equation}
It is essential for our discussion that the energy dependence of the diffusion coefficient changes at a critical energy $E_{0,i}= eZ_i E_0 $ which is different for different nuclei.

For $l_0 \sim 0.3$~kpc, the mean time of diffusion of nuclei in the galaxy is
\begin{equation}
t_D\sim \frac{R^2}{D} \sim 10^7{\rm yr}\left (\frac{R}{10\ {\rm kpc}} \right )^2 \left (\frac{26}{Z} \ \frac{10^{19} \, {\rm eV}}{E}
\right )^{2-\delta_2},
\label{t_D}
\end{equation}
where $R\sim 10$~kpc is the diffusion distance scale.  For distances of the order of the Sun's Galactocentric distance, $R\sim 8$~kpc,
this escape time is much longer than the time between individual GRBs, $ t_D \gg t_{GRB}\sim 10^4-10^5$~yr, so the UHECR injection reaches a steady-state regime.

We are interested in a steady-state solution, so that $n_i(E,\vec{r},t)$ should not depend on time.  However, a few latest GRBs can cause fluctuations on the average, as discussed below.  Let us first assume that Galactic halo is a sphere with radius $R_G\sim 100$~kpc and that all the sources are at the Galactic Center, so that the problem is spherically symmetric: $Q_i(E, \vec{r}, t)= \delta(\vec{r} ) Q_0(E_0/E)^\gamma$,
 $n_i(E,\vec{r},t)=n_i(E,r)$. This is admittedly a simplified model, and we will replace it with a more realistic model below.
Neglecting the energy losses inside the Galaxy, one obtains the solution of Eq.~(\ref{transport}) with a boundary condition corresponding to a diminishing flux outside the Galaxy:
\begin{equation}
n_i(E,r) = \frac{Q_0}{4\pi r\, D_i(E)} \left ( \frac{E_0}{E} \right )^\gamma.
\label{solution_GC}
\end{equation}

This solution corresponds to energy-dependent composition for $E>E_0$.  Indeed, at critical energy $E_{0,i}$, which is different for each nucleus, the solution (\ref{solution_GC}) changes from $\propto E^{-\gamma-\delta_1} $ to $\propto E^{-\gamma-2+\delta_2} $ because of the change in $D_i(E)$.  Since the change occurs at a rigidity-dependent critical energy $E_{0,i}=e E_0 Z_i$, the larger nuclei lag behind the lighter nuclei in terms of the critical energy and the change in slope.  If protons dominate for $E<E_0$, their flux drops dramatically for $E>E_0$, and the heavier nuclei dominate the flux.  The higher $Z_i$, the higher is the energy at which the species experiences a drop in flux.

One can also understand the change in composition by considering the time of diffusion across the halo is $t_i \sim R^2/D_i$.  The longer the particle remains in the halo, the higher is the probability of its detection.   At higher energies, the magnetic field's ability to delay the passage of the particle diminishes, and the density of such particles drops precipitously for $E>E_{0,i}$.  Since $E_i$ is proportional to the  electric charge, the drop in the flux occurs at different energies for different species.

Of course, the assumption that all the sources are located in the Galactic center may not be realistic.
If past GRBs in the Milky Way are the sources, one can model their distribution in different ways: one can assume (i) that all GRBs happen in the Galactic Center, or (ii) that GRB distribution follows the distribution of stars in MW, or (iii) one can include the short GRB distribution, which is expected to extend more into the halo.

In Fig. 1 we show the spectrum calculated numerically for the source distribution (ii), which we model using the star counts from Ref.~\cite{Bahcall}.  Some $10^3$ GRBs separated by time intervals of $10^5$ years were generated in each Monte Carlo simulation, and the parameters were chosen to fit the data.  We have assumed a two-component composition with protons and iron nuclei.   The best fit for $\gamma=2.3$ is obtained for 90\% protons and 10\% iron, and  $\approx 4\mu$G magnetic field  coherent on 0.2~kpc scale.  For the case of short GRBs (iii), the distribution of sources can be obtained from observations~\cite{shortGRBs}.  The spectra obtained in this case are similar to those shown in Fig.~1.

It is intriguing that the change from proton to iron in Fig.~1 is consistent with the dip in the spectrum that is usually attributed to either pair production or the change from Galactic to extragalactic component~\cite{Hill:1983mk,Berezinsky:2004wx}.  However, one should not consider the fit in Fig.~1 more than an illustration of the general principle.  One must include multiple species of nuclei and the extragalactic protons, and one must model the propagation of UHECRs more carefully to compare the predictions with the data~\cite{Abraham:2009dsa} quantitatively.

The data do not show a significant Galactocentric anisotropy in the arrival directions of UHECRs (although some clusters reported by PAO tend to gravitate toward the Galactic plane). For nuclei, however, one does not expect much anisotropy even if the sources are Galactic. We define the anisotropy parameter  $\delta (E)$ in terms of maximal and minimal fluxes $J_{\rm min}(E)$ and
$J_{\rm max}(E)$, depending on the arrival directions.  In the diffusion approximation,
\begin{equation}
\delta (E)\equiv \frac{J_{\rm max}-J_{\rm min}}{J_{\rm max}+J_{\rm min}}=
3 D(E) \, \frac{\partial}{\partial r} \ln \sum_i n_i(E,r).
\end{equation}
Obviously, model (i), assuming that all the sources are in the Galactic Center, predicts the largest anisotropy, hence setting the upper bound on $\delta$.  We find $\delta<0.1$ for $E<3\times 10^{19}$~eV.

For model (ii), which assumes that the source distribution follows the stellar distribution, the anisotropy can be computed numerically.  We have calculated the anisotropy parameter by generating $10^3$ GRBs occurring once every $10^5$~years.
The results are shown in Fig.~2 for the same parameters as in Fig.~1.  The anisotropy for model (iii) is even smaller.

While the average flux includes contributions of GRBs form different distances and different times, the latest nearby GRBs can create fluctuations.  A cluster of several UHECRs around Cen A detected by PAO may be the result of such a fluctuation due to one  GRB that happens to coincide with Cen A.  Alternatively, since we expect the high-energy protons to escape from our Galaxy and from other galaxies, the cluster around Cen A may be due to extragalactic protons. Unlike protons, UHE nuclei from the last GRB do not introduce a large degree of anisotropy, as one can see form Fig.~2 based on a semi-realistic Monte Carlo simulation.

Our model can be improved.  First, one can use a more realistic source population model.  Second, one should include the coherent component of the Galactic magnetic field.  Third, one should not assume that UHECRs comprise only two types of particles, and one should include a realistic distribution of nuclei. Finally, one should include the extragalactic component of UHECRs produced by distant sources, such as active galactic nuclei (AGN) and GRBs (outside MW).  A recent realization that very high energy gamma rays observed by Cherenkov telescopes from distant blazars are likely to be secondary photons produced in cosmic ray interactions along the line of sight lends further support to the assumption that cosmic rays are copiously produced in AGN jets~\cite{Essey:2009zg}. For energies $E>3\times 10^{19}$~eV, the energy losses due to photodisintegration, pion production, pair production and interactions with interstellar medium become important and must be included. The propagation distance in the Galaxy exceeds 10~Mpc, according to Eq.~(\ref{t_D}), so that the Galactic component should exhibit an analog of GZK suppression in the spectrum.    The extragalactic propagation can also affect the composition around $10^{18}$~eV~\cite{Hill:1983mk}, as well as the arrival direction anisotropy~\cite{propagation}.

We have shown that effects of rigidity-dependent diffusion of UHECRs from possible Galactic sources, such as past GRBs (or rare types of supernovae), in the Milky Way can produce the energy-dependent composition as observed by PAO~\cite{Abraham:2009dsa}.  The simplest two-component model including protons and iron nuclei from the Galactic sources gives a good fit to composition~\cite{Abraham:2009dsa} and spectrum~\cite{Abraham:2009wk} for reasonable values of the Galactic magnetic fields.

A.K. was supported  by DOE grant DE-FG03-91ER40662 and NASA ATFP grant  NNX08AL48G. S.N. was supported by Grant-in-Aid for Scientific Research No.19047004, No.21105509, and the Global COE Program by Ministry of Education, Culture, Sports, Science and Technology (MEXT) in Japan and Grant-in-Aid No.19104006, No.19740139, and No.21540304 by Japan Society for the Promotion of Science (JSPS). A. K. thanks Aspen Center for Physics for hospitality. 


\end{document}